\documentclass{article}
\usepackage{graphicx} 
\usepackage{amsmath,amssymb}
\usepackage{diagbox}
\usepackage{algorithm}
\usepackage{algorithmic}

\usepackage{authblk}
\usepackage{hyperref} 

\author[1]{Jiawei Nie\thanks{Email: \href{mailto:21377128@buaa.edu.cn}{21377128@buaa.edu.cn}}}
\author[1]{Yongjun Wang\thanks{Email: \href{mailto:wangyj@buaa.edu.cn}{wangyj@buaa.edu.cn}}}
\author[1]{Songyi Liu\thanks{Email: \href{mailto:liusongyi@buaa.edu.cn}{liusongyi@buaa.edu.cn}}}
 
\affil[1]{School of Mathematical Sciences, Beihang University, Beijing 100191, China}

\usepackage[backend=biber,sorting=none,url=false,doi=false,isbn=false,date=year]{biblatex}

\begin{document}

\title{Determining Strong Contextuality on rank-one Projectors}
\date{}
\maketitle

\abstract{The strength of quantum contextuality is closely related to quantum computation power. Yu-Oh set is the minimal quantum system with state-independent contextuality(SIC). However, its strength of the contextuality has not been taken into account. In this paper, we present a general method to determine whether there is a quantum state with strong contextuality in the quantum system composed of rank-one projectors. Based on this method, we conclude that Yu-Oh set does not have quantum states with strong contextuality. This indicates that strong contextuality and SIC are mutually independent.  }

\section{Introduction}
Quantum contextuality, as the most fundamental non-classical feature of quantum systems, has been identified as a source of quantum computational speedups and quantum advantage in quantum computing\cite{howardContextualitySuppliesMagic2014,raussendorfContextualityMeasurementbasedQuantum2013}, significantly influencing the computational power of quantum computing. This concept describes that measurements of quantum observables cannot be considered as revealing pre-existing values. Bell was the first to propose a rigorous mathematical formalism, namely inequalities, to reveal this property of quantum systems\cite{bellProblemHiddenVariables1966}, but his focus was on spatially separated systems. Subsequently, Kochen and Specker unveiled the contextuality underlying general quantum systems by constructing diagrams\cite{kochenProblemHiddenVariables1975}. This marked the origin of the general concept of contextuality. Since then, researchers have been further classifying and refining the concept of contextuality, and have proposed numerous mathematical frameworks to describe it. Abramsky and Brandenburger, in their proposed sheaf-theoretical framework, have stratified the strength of contextuality\cite{abramskySheaftheoreticStructureNonlocality2011}. Contextuality, based on its strength, is classified from strong to weak as strong contextuality, logical contextuality, and probabilistic contextuality. The stronger the contextuality, the greater the computational power of the quantum system. Therefore, determining the strength of contextuality of quantum systems is highly significant for the study of quantum computing applications. Also, based on whether quantum systems exhibit contextuality under different quantum states, contextuality can be further divided into state-independent contextuality and state-dependent contextuality. However, the relationship between contextuality under these two different classification standards remains unclear. The original Kochen-Specker proof of contextuality serves as both an example of strong contextuality and state-independent contextuality\cite{kochenProblemHiddenVariables1975}. Greenberger–Horne–Zeilinger proof of contextuality serves as an example of strong contextuality, but it is not an example of state-independent contextuality\cite{greenbergerGoingBellsTheorem2007}. However, there has yet to be an example of state-independent contextuality that is not also strongly contextual. The Yu-Oh set\cite{yuStateIndependentProofKochenSpecker2012}, as the smallest quantum system with state-independent contextuality\cite{cabelloQuantumStateindependentContextuality2016}, has not yet been systematically studied for the strength of its contextuality. In this paper, we propose a general method to determine whether a quantum system composed of rank-one projectors exhibits strong contextuality under a certain quantum state. When examining the Yu-Oh set, we conducted a further analysis of the method and ultimately proved that the Yu-Oh set does not exhibit strong contextuality under any quantum state, which also demonstrates the mutual independence between state-independent contextuality and strong contextuality. The organization of this paper is as follows: In Section 2, we introduce the mathematical framework related to strong contextuality; in Section 3, we propose a general algorithm to determine the existence of strong contextuality states; in Section 4, we further analyze this algorithm and apply it to the Yu-Oh set; and in Section 5, we present our conclusions.

\section{Background}
In the Abramsky-Brandenburger sheaf-theoretic framework, quantum systems are described using measurement scenarios $(\mathcal{M, C})$. $\mathcal{M}$ is the set of all measurements, and $\mathcal{C}$ is the set of all contexts. A context is a set $C \subset \mathcal{M}$ representing a maximal set of compatible (i.e., comeasurable) measurements. A measurement gets a value from the outcome set $\mathcal{O}$(usually $\{0,1\}$). In a quantum system composed of rank-one projectors, each projector $P$ corresponds to a measurement $M$, and in Hilbert space, each projector corresponds to a closed linear subspace. Measurements that are co-measurable correspond to mutually orthogonal projectors. All projectors within a context correspond to an orthogonal set in Hilbert space. In an experiment, one and only one projector among a set of orthogonal bases will get the value 1. Therefore, when the orthogonal set corresponding to the projectors in a context is incomplete, we add the projector corresponding to the orthogonal complement of this set to the context. If the original projectors all get the value 0, the newly added projector gets the value 1. An event is a mapping from a certain context $C$ to outcome set $\mathcal{O}$. When the state $|\psi\rangle$ of the system is given, the probability of a certain projector $P$ getting the value 1 is $\langle\psi|P|\psi\rangle$. If the probability is greater than 0, the corresponding event is possible. An empirical model is an event probability table corresponding to the quantum state. A global assignment $v$ is a mapping from $\mathcal{M}$ to $\mathcal{O}$. If, for a given quantum state, there does not exist a global assignment such that its restriction in every context is a possible event, then the quantum system exhibits strong contextuality under that state.

Now we give an example of a bipartite Bell-type scenario, where Alice has two measurements $\{a,a'\}$, and Bob has $\{b,b'\}$. There are two outcomes, $0$ or $1$, for each measurement. $\mathcal{M}=\{a,a',b,b'\}$, $\mathcal{C}=\{\{a,b\},\{a,b'\},\{a',b\},\{a',b'\}\}$ and $\mathcal{O}=\{0,1\}$. When the quantum state of the system is given, we can get the corresponding empirical model as follows: 

\begin{table}[h!]
    \centering
    \begin{tabular}{|c|c|c|c|c|}
        \hline
        \diagbox{context}{outcome}  & 0 0 & 0 1 & 1 0 & 1 1 \\ \hline
        $a\quad b$       & $p_1$        & $p_2$ & $p_3$    & $p_4$    \\ \hline
        $a\quad b'$       & $p_5$        & $p_6$ & $p_7$    & $p_8$     \\ \hline
        $a'\quad b$       & $p_9$        & $p_{10}$ & $p_{11}$    & $p_{12}$     \\ \hline
        $a'\quad b'$       & $p_{13}$        & $p_{14}$ & $p_{15}$    & $p_{16}$     \\ \hline
    \end{tabular}
    \caption{empirical model for the bipartite Bell-type scenario}
\end{table}

Let us take the Yu-Oh set as an example. The Yu-Oh set consists of 13 rays. Each ray $|\psi\rangle$ corresponds to a rank-one projector $P=|\psi\rangle\langle\psi|$, and mutually orthogonal rays form a context. We complete the contexts containing only two rays. Since the newly added ray is only intended to represent the value 1 for the first two rays taking the value 0, we do not consider the new orthogonality relations that it may introduce. After completing the Yu-Oh set, all the rays are as follows, with $1'$ to $12'$ being the newly added rays:

\[
|1\rangle = (1, 0, 0) \quad |2\rangle = (0, 1, 0) \quad |3\rangle = (0, 0, 1)
\]
\[
|4\rangle = (0, 1, -1) \quad |5\rangle = (1, 0, -1) \quad |6\rangle = (1, -1, 0)
\]
\[
|7\rangle = (0, 1, 1) \quad |8\rangle = (1, 0, 1) \quad |9\rangle = (1, 1, 0)
\]
\[
|A\rangle = (-1, 1, 1) \quad |B\rangle = (1, -1, 1) \quad |C\rangle = (1, 1, -1) \quad |D\rangle = (1, 1, 1)
\]
\[
|1'\rangle = (2, 1, 1) \quad |2'\rangle = (-1, -2, 1) \quad |3'\rangle = (1, -1, 2)
\]
\[
|4'\rangle = (-1, -2, -1) \quad |5'\rangle = (2, 1, -1) \quad |6'\rangle = (1, -1, -2)
\]
\[
|7'\rangle = (1, 1, 2) \quad |8'\rangle = (-2, 1, -1) \quad |9'\rangle = (-1, 2, 1)
\]
\[
|10'\rangle = (2, -1, -1) \quad |11'\rangle = (1, -2, 1) \quad |12'\rangle = (-1, -1, 2)
\]

Therefore, the empirical model of completed Yu-Oh set is presented in the following table, where $*$ denotes the probability is uncertain:

\begin{table}[h!]
    \centering
    \begin{tabular}{|c|c|c|c|}
        \hline
        \diagbox{context}{outcome}  & 1 0 0 & 0 1 0 & 0 0 1 \\ \hline
        $1\quad 2\quad 3$       & *        & * & *        \\ \hline
        $1\quad 4\quad 7$       & *        & * & *         \\ \hline
        $2\quad 5\quad 8$       & *        & * & *         \\ \hline
        $3\quad 6\quad 9$       & *        & * & *         \\ \hline
        $4\quad A\quad 1'$       & *        & * & *         \\ \hline
        $8 \quad A\quad 2'$      & *        & * & *         \\ \hline
        $9\quad A\quad 3'$       & *        & * & *         \\ \hline
        $5\quad B\quad 4'$       & *        & * & *         \\ \hline
        $7\quad B\quad 5'$       & *        & * & *         \\ \hline
        $9\quad B\quad 6'$       & *        & * & *         \\ \hline
        $6\quad C\quad 7'$       & *        & * & *         \\ \hline
        $7\quad C\quad 8'$       & *        & * & *         \\ \hline
        $8\quad C\quad 9'$       & *        & * & *         \\ \hline
        $4\quad D\quad 10'$       & *        & * & *         \\ \hline
        $5\quad D\quad 11'$       & *        & * & *         \\ \hline
        $6\quad D\quad 12'$       & *        & * & *         \\ \hline
    \end{tabular}
    \caption{empirical model for completed Yu-Oh set}
\end{table}

For quantum systems composed of real projectors, the existence of strongly contextual quantum states can be reduced to the existence of strongly contextual real quantum pure states. The following two lemmas illustrate this point.

{\em Lemma~1:} If a quantum system does not exhibit strong contextuality in any pure state, then it does not exhibit strong contextuality in any quantum state.

{\em Proof:} If a quantum system does not exhibit strong contextuality in any pure state, then for every pure state there exists a global assignment $s$ whose restriction in every context is a possible event. For any mixed state $\rho = \sum\limits_i p_i |\psi_i \rangle \langle \psi_i|$, there is a global assignment $s$ whose restriction in every context is a possible event in the pure state $|\psi_1\rangle$. Then for every projector $P$, $\langle\psi_1|P|\psi_1\rangle>0$. Then for the mixed state $\rho$, the probability of $P$ getting the value 1 is $tr(\rho P)=\sum\limits_i p_i\langle\psi_i|P|\psi_i\rangle>0$. 

{\em Lemma~2:} Assuming that all projectors are real, if a quantum system does not exhibit strong contextuality in any real pure state, then it does not exhibit strong contextuality in any pure state.

{\em Proof:} For any pure state $|\psi\rangle=|\psi_1\rangle+i|\psi_2\rangle$, here $|\psi_1\rangle$ and $|\psi_2\rangle$ are real, if a quantum system does not exhibit strong contextuality in any real pure state, then for $|\psi_1\rangle$, there exists a global assignment $s$ whose restriction in every context is a possible event. Then for every projector $P$, we have $\langle\psi_1|P|\psi_1\rangle>0$. Then for the pure state $|\psi\rangle$, the probability of $P$ getting the value 1 is $\langle\psi|P|\psi\rangle=(\langle\psi_1|-i\langle\psi_2|)P(|\psi_1\rangle+i|\psi_2\rangle)=\langle\psi_1|P|\psi_1\rangle+\langle\psi_2|P|\psi_2\rangle-i\langle\psi_2|P|\psi_1\rangle+i\langle\psi_1|P|\psi_2\rangle=\langle\psi_1|P|\psi_1\rangle+\langle\psi_2|P|\psi_2\rangle>0$. Here $-i\langle\psi_2|P|\psi_1\rangle+i\langle\psi_1|P|\psi_2\rangle=0$, because $|\psi_1\rangle,|\psi_2\rangle,P$ are real.

\section{Method}

For an $n$ dimensional quantum system composed of real rank-one projectors, if a context is not composed of $n$ rank-one projectors, we need to complete it. In an experiment, there is exactly one projector in the orthogonal basis that takes the value of 1, i.e., it is assigned the value of 1. We only consider such global assignments. Since the existence of strongly contextual quantum states in a quantum system composed of real projectors can be reduced to the existence of real pure states, we only consider real pure states.

Let $A$ be the set of all global assignments that satisfy the aforementioned requirements. For a given global assignment $v_i$, $E_i$ denotes the set of all real pure states for which there exists a context where the restriction of the global assignment in that context is not a possible event. Then, the set of strongly contextual real pure states is $\bigcap\limits_{v_i\in A}E_i$. For a given context $C_j$ and a given global assignment $v_i$, let $F_j^i$ be the set of all real pure states such that the restriction of the global assignment in this context is a possible event. Then, $E_i=\overline{\bigcap\limits_{C_j\in \mathcal{C}}F_j^i}=\bigcup\limits_{C_j\in \mathcal{C}}\overline{F_j^i}$. In a quantum system composed of real projectors, $\overline{F_j^i}$ is the orthogonal complement of the projector $P$ assigned the value 1 in a given context $C_j$, that is, $P^{\perp}$. Based on orthogonality relations, $P^{\perp}$ is composed of other projectors in the context $C_j$ that are assigned the value 0. Let $B_i=\{P\in\mathcal{M}|v_i(P)=1\}$, we have $E_i=\bigcup\limits_{P\in B_i}P^{\perp}$.

For determining the existence of strongly contextual quantum states in a quantum system composed of real rank-one projectors, we propose the following algorithm:

\begin{algorithm}
    \renewcommand{\algorithmicrequire}{\textbf{Input:}}	\renewcommand{\algorithmicensure}{\textbf{Output:}}
    \caption{Determining Strong Contextuality of General Systems}
    \begin{algorithmic}[1]
    \REQUIRE projectors $P$ which compose of the completed quantum system.
    \STATE Use some algorithm(such as branch and bound) to determine $A$.
    \STATE Calculate the orthogonal complement $P^{\perp}$ of every projector $P$ evaluated to 1 in each global assignment $v_i$
    \STATE Check whether the set $\bigcap\limits_{v_i\in A}\bigcup\limits_{P\in B_i}P^{\perp}$ is empty.
    \IF{empty} \STATE determine that the system is not strongly contextual
    \ELSE   \STATE the set $\bigcap\limits_{v_i\in A}\bigcup\limits_{P\in B_i}P^{\perp}$ is all real pure states with strong contextuality
    \ENDIF
    \ENSURE real pure states with strong contextuality if any.
    \end{algorithmic}
\end{algorithm}

Let us analyze the complexity of the algorithm without considering the time taken to determine $A$. The number of projectors assigned 1 in each global assignment is no more than the number of contexts $C$. The number of all global assignments is no more than $d^C$, where $d$ denotes the dimension of the quantum system. The time complexity of the algorithm is $O(Cd^C)$.

\section{Yu-Oh Set}

For the Yu-Oh set, we conduct further analysis based on the original method. For any projector $P$, if there are two global assignments $v, v'$ such that $v(P)=1, v'(P)=0$, then $E$ contains $P^{\perp}$, while $E'$ does not contain $P^{\perp}$, and we only need to consider whether the intersection of $P^{\perp}$ and the orthogonal complements of the other projectors lies within $\bigcap\limits_{v_i\in A}\bigcup\limits_{P\in B_i}P^{\perp}$. 

 And now we give all global assignments in the Yu-Oh set as follows:

 \begin{table}[h!]
    \centering
    \begin{tabular}{|c|c|c|c|c|c|c|c|c|c|c|c|c|c|}
        \hline
        \diagbox{index}{projector}  & $1$ & $2$ & $3$ & $4$ & $5$ & $6$ & $7$ & $8$ & $9$ & $A$ & $B$ & $C$ & $D$  \\ \hline
        $1$       & 0        & 0 & 1 & 0 &0 & 0 &1&1&0&0&0&0&0       \\ \hline
        $2$       & 0        & 0 & 1 & 0 & 1 & 0 & 1 & 0 & 0 & 0 & 0 & 0 & 0       \\ \hline
        $3$       & 0  & 0 & 1 & 0 & 0 & 0 & 1 & 1 & 0 & 0 & 0 & 0 & 1         \\ \hline
        $4$       & 0  & 0 & 1 & 1 & 0 & 0 & 0 & 1 & 0 & 0 & 0 & 0 & 0 \\ \hline
        $5$       & 0  & 0 & 1 & 1 & 1 & 0 & 0 & 0 & 0 & 0 & 0 & 0 & 0        \\ \hline
        $6$       & 0  & 0 & 1 & 1 & 1 & 0 & 0 & 0 & 0 & 0 & 0 & 1 & 0         \\ \hline
        $7$       & 0  & 0 & 1 & 1 & 0 & 0 & 0 & 1 & 0 & 0 & 1 & 0 & 0     \\ \hline
        $8$       & 0  & 1 & 0 & 0 & 0 & 0 & 1 & 0 & 1 & 0 & 0 & 0 & 0      \\ \hline
        $9$       & 0  & 1 & 0 & 0 & 0 & 1 & 1 & 0 & 0 & 0 & 0 & 0 & 0     \\ \hline
        $10$      & 0  & 1 & 0 & 0 & 0 & 0 & 1 & 0 & 1 & 0 & 0 & 0 & 1       \\ \hline
        $11$      & 0  & 1 & 0 & 1 & 0 & 0 & 0 & 0 & 1 & 0 & 0 & 0 & 0          \\ \hline
        $12$      & 0  & 1 & 0 & 1 & 0 & 1 & 0 & 0 & 0 & 0 & 0 & 0 & 0       \\ \hline
        $13$      & 0  & 1 & 0 & 1 & 0 & 0 & 0 & 0 & 1 & 0 & 0 & 1 & 0     \\ \hline
        $14$      & 0  & 1 & 0 & 1 & 0 & 1 & 0 & 0 & 0 & 0 & 1 & 0 & 0         \\ \hline
        $15$      & 0  & 0 & 1 & 0 & 1 & 0 & 1 & 0 & 0 & 1 & 0 & 0 & 0         \\ \hline
        $16$      & 0  & 1 & 0 & 0 & 0 & 1 & 1 & 0 & 0 & 1 & 0 & 0 & 0        \\ \hline
        $17$      & 1  & 0 & 0 & 0 & 0 & 0 & 0 & 1 & 1 & 0 & 0 & 0 & 0  \\ \hline
        $18$      & 1  & 0 & 0 & 0 & 0 & 1 & 0 & 1 & 0 & 0 & 0 & 0 & 0     \\ \hline
        $19$      & 1  & 0 & 0 & 0 & 1 & 0 & 0 & 0 & 1 & 0 & 0 & 0 & 0     \\ \hline
        $20$      & 1  & 0 & 0 & 0 & 1 & 1 & 0 & 0 & 0 & 0 & 0 & 0 & 0     \\ \hline
        $21$      & 1  & 0 & 0 & 0 & 0 & 0 & 0 & 1 & 1 & 0 & 0 & 0 & 1         \\ \hline
        $22$      & 1  & 0 & 0 & 0 & 1 & 0 & 0 & 0 & 1 & 0 & 0 & 1 & 0     \\ \hline
        $23$      & 1  & 0 & 0 & 0 & 0 & 1 & 0 & 1 & 0 & 0 & 1 & 0 & 0     \\ \hline
        $24$      & 1  & 0 & 0 & 0 & 1 & 1 & 0 & 0 & 0 & 1 & 0 & 0 & 0    \\ \hline
    \end{tabular}
    \caption{global assignments for completed Yu-Oh set}
\end{table}

\begin{table}[h!]
    \centering
    \begin{tabular}{|c|c|c|c|c|c|c|c|c|c|c|c|c|}
    \hline
     \diagbox{index}{projector} & $1'$ & $2'$ & $3'$ & $4'$ & $5'$ & $6'$ & $7'$ & $8'$ & $9'$ & $10'$ & $11'$ & $12'$ \\ \hline
     $1$ &1&0&1&1&0&1&1&0&0&1&1&1 \\ \hline
     $2$ & 1 & 1 & 1 & 0 & 0 & 1 & 1 & 0 & 1 & 1 & 0 & 1 \\ \hline
     $3$ & 1 & 0 & 1 & 1 & 0 & 1 & 1 & 0 & 0 & 0 & 0 & 0 \\ \hline
     $4$ & 0 & 0 & 1 & 1 & 1 & 1 & 1 & 1  & 0 & 0 & 1 & 1   \\ \hline     
     $5$ & 0 & 1 & 1 & 0 & 1 & 1 & 1 & 1 & 1 & 0 & 0 & 1 \\ \hline
     $6$ & 0 & 1 & 1 & 0 & 1 & 1 & 0 & 0 & 0 & 0 & 0 & 1 \\ \hline
     $7$ & 0 & 0 & 1 & 0 & 0 & 0 & 1 & 1 & 0 & 0 & 1 & 1     \\ \hline
     $8$ & 1 & 1 & 0 & 1 & 0 & 0 & 1 & 0 & 1 & 1 & 1 & 1  \\ \hline
     $9$ & 1 & 1 & 1 & 1 & 0 & 1 & 0 & 0 & 1 & 1 & 1 & 0     \\ \hline
     $10$ & 1 & 1 & 0 & 1 & 0 & 0 & 1 & 0 & 1 & 0 & 0 & 0   \\ \hline
     $11$ & 0 & 1 & 0 & 1 & 1 & 0 & 1 & 1 & 1 & 0 & 1 & 1\\ \hline
     $12$  & 0 & 1 & 1 & 1 & 1 & 1 & 0 & 1 & 1 & 0 & 1 & 0  \\ \hline
     $13$ & 0 & 1 & 0 & 1 & 1 & 0 & 0 & 0 & 0 & 0 & 1 & 1     \\ \hline
     $14$ & 0 & 1 & 1 & 0 & 0 & 0 & 0 & 1 & 1 & 0 & 1 & 0 \\ \hline
     $15$ & 0 & 0 & 0 & 0 & 0 & 1 & 1 & 0 & 1 & 1 & 0 & 1 \\ \hline
     $16$ & 0 & 0 & 0 & 1 & 0 & 1  & 0 & 0 & 1 & 1 & 1 & 0 \\ \hline
     $17$  & 1 & 0 & 0 & 1 & 1 & 0 & 1 & 1 & 0 & 1 & 1 & 1     \\ \hline
     $18$ & 1 & 0 & 1 & 1 & 1 & 1 & 0 & 1 & 0 & 1 & 1 & 0 \\ \hline
     $19$ & 1 & 1 & 0 & 0 & 1 & 0 & 1 & 1 & 1 & 1 & 0 & 1  \\ \hline
     $20$ & 1 & 1 & 1 & 0 & 1 & 1 & 0 & 1 & 1 & 1 & 0 & 0   \\ \hline
     $21$ & 1 & 0 & 0 & 1 & 1 & 0 & 1 & 1 & 0 & 0 & 0 & 0  \\ \hline
     $22$ & 1 & 1 & 0 & 0 & 1 & 0 & 0 & 0 & 0 & 1 & 0 & 1  \\ \hline
     $23$ & 1 & 0 & 1 & 0 & 0 & 0 & 0 & 1 & 0 & 1 & 1 & 0  \\ \hline
     $24$  & 0 & 0 & 0 & 0 & 1 & 1 & 0 & 1 & 1 & 1 & 0 & 0   \\ \hline
    \end{tabular}
    \caption{global assignments for completed Yu-Oh set(continued table)}
\end{table}

We find that for any projector $P$ in the Yu-Oh set, there are two global assignments $v, v'$ such that $v(P)=1,v'(P)=0$. So we only need to consider the intersection of $P^{\perp}$.

Suppose that two projectors are $P_1$ and $P_2$, the rays corresponding to them are $|\psi_1\rangle,|\psi_2\rangle$, and we denote the intersection of $P_1^{\perp},P_2^{\perp}$ by $I$. Suppose another projector is $P$, the ray corresponding to it is $|\psi\rangle$. If there exists a vector in $I$ that is orthogonal to $|\psi\rangle$, and since the vectors in $I$ are orthogonal to $|\psi_1\rangle,|\psi_2\rangle$, we know that $|\psi\rangle$ lies within the plane spanned by $|\psi_1\rangle,|\psi_2\rangle$ in a three dimensional Hilbert space. Therefore, we need to determine whether a global assignment exists that assigns the value 1 to a ray, which is orthogonal to both $|\psi_1\rangle$ and $|\psi_2\rangle$, or assigns the value 0 to all rays within the plane spanned by $|\psi_1\rangle$ and $|\psi_2\rangle$. If such an assignment exists, it implies that $I$ is not in $\bigcap\limits_{v_i\in A}\bigcup\limits_{P\in B_i}P^{\perp}$.

We can divide any two rays $|\psi_1\rangle,|\psi_2\rangle$ into the following three situations:

\begin{enumerate}
    \item $|\psi_1\rangle,|\psi_2\rangle$ are orthogonal, such as $|1\rangle,|2\rangle$. Then $|\psi_1\rangle,|\psi_2\rangle$ must be included in a context $C_j$. Assuming $C_j=\{|\psi_1\rangle,|\psi_2\rangle,|\psi_3\rangle\}$, there is a global assignment $v$ that satisfies $v(|\psi_3\rangle)=1$. Therefore, $I$ corresponding to $|\psi_1\rangle,|\psi_2\rangle$ is not in $\bigcap\limits_{v_i\in A}\bigcup\limits_{P\in B_i}P^{\perp}$. 
    \item $|\psi_1\rangle,|\psi_2\rangle$ are not orthogonal, but the contexts containing them have the same ray, such as $|1\rangle,|5\rangle$. $\{|1\rangle,|2\rangle,|3\rangle\}$ and $\{|2\rangle,|5\rangle,|8\rangle\}$ both contain $|2\rangle$, so $|2\rangle$ is orthogonal to $|1\rangle$ and $|5\rangle$. Since there is a global assignment $v$ that assigns $|2\rangle$ to 1, we can be sure that the intersection $I$ corresponding to $|1\rangle$ and $|5\rangle$ is not in $\bigcap\limits_{v_i\in A}\bigcup\limits_{P\in B_i}P^{\perp}$. 
    \item $|\psi_1\rangle,|\psi_2\rangle$ are not orthogonal, and there is no identical ray in the contexts containing them. We need to calculate all the rays that can be linearly represented by $|\psi_1\rangle,|\psi_2\rangle$, and then determine whether a global assignment $v$ exists that assigns these rays to 0.
\end{enumerate}

We find that the first and second situations cover any two rays between $|1\rangle\sim|D\rangle$, namely, the original Yu-Oh set. The newly added rays and the other rays primarily fall into the third situation. For the third situation, we use a Python program to calculate rays and find a global assignment that satisfies the conditions. Our final conclusion is that $\bigcap\limits_{v_i\in A}\bigcup\limits_{P\in B_i}P^{\perp}$ is empty, and the Yu-Oh set does not exhibit strong contextuality in any real pure state, implying that it does not exhibit strong contextuality in any quantum state.

\section{Conclusion}

We propose a general method to determine the existence of strongly contextual quantum states in quantum systems composed of real rank-one projectors, and conclude that the Yu-Oh set does not exhibit strong contextuality in any quantum state. This is beneficial for assessing the computational capabilities of quantum systems and also reveals the mutual independence of strong contextuality and state-independent contextuality. We can consider utilizing mathematical tools such as graph theory to conduct further complexity analysis of the algorithm. Additionally, we can also explore extending this method to quantum systems composed of high-rank projectors, or generalizing the approach tailored for Yu-Oh sets.

\printbibliography

\end{document}